\documentclass[aps,prl,twocolumn]{revtex4}
\usepackage{amsmath,bm,epsfig,,subfigure}
\usepackage{color}
\usepackage[normalem]{ulem}

\def\Fbox#1{\vskip1ex\hbox to 8.5cm{\hfil\fboxsep0.3cm\fbox{%
  \parbox{8.0cm}{#1}}\hfil}\vskip1ex\noindent}  %%  {TEXT} in BOX

\newcommand{\B}[1]{{\bm{#1}}}%% Bold Roman & Greek Lower & Upper Case
\newcommand{\C}[1]{{\mathcal{#1}}}    %%   Calligrapfic Upper case
\let \= \equiv \let\*\cdot \let\~\widetilde \let\-\overline

\begin{document}
\title{Microscopic Mechanism of Shear Bands in Amorphous Solids}
\author{Ratul Dasgupta, H. George E. Hentschel and  Itamar Procaccia}
\affiliation{Department of Chemical Physics, The Weizmann
 Institute of Science}
\date{\today}
\begin{abstract}
The fundamental instability responsible for the shear localization which results in shear bands in amorphous solids remains unknown despite enormous amount of research, both experimental and theoretical. As this is the main mechanism for the failure of metallic glasses, understanding the instability is
invaluable in finding how to stabilize such materials against the tendency to shear localize. In this Letter we explain the mechanism for shear localization under shear, which is the appearance of highly correlated lines of Eshelby-like quadrupolar singularities which organize the non-affine plastic flow of the amorphous solid into a shear band. We prove analytically that such highly correlated solutions in which $\C N$ quadrupoles are aligned with equal orientations are minimum energy states when the strain is high enough. The line lies at 45 degrees to the compressive stress.
%This result sounds the death knell to all theories of plasticity that assume statistical independence of plastic events %(shear transformation zones).

\end{abstract}

\maketitle

\begin{figure}
\includegraphics[scale = 0.30]{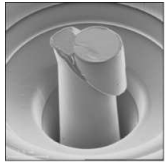}
\caption{A typical example of the failure of a metallic glass sample when subjected to compressive stress. The material localizes the stress in a plane that is at 45 degrees to the compressive stress axis, and then breaks along this plane \cite{08Che}.}
\label{shearband}
\end{figure}

In Fig.\!\!~\ref{shearband} we show the typical failure of a sample of metallic glass when subjected
to compressive stress. When the stress exceeds some critical level (known as the {\em yield stress}), the sample, rather than flowing
%%%%%%%%%%%%%%%%%%%%%%%%%%%%%%%%%%%%%%%%%%%%%%%%%%
%%%%%%%%%%%%%%%%%%%%%%%%%%%%%%%%%%%%%%%%%%%%%%%%%
homogeneously in a plastic flow, localizes all the shear in a plane that is at 45 degrees to the compressive stress axis, and then breaks along this plane \cite{08Che}. Despite considerable amount
of research \cite{82SSH,02HFOLB,07SKLF,06TLB}, the precise mechanism that gives rise to this spectacular phenomenon remains elusive. In this Letter we close this gap.

Amorphous solids are obtained when a glass-former is cooled below the glass transition \cite{08Che,06Dyr,09Cav}
to a state which on the one hand is amorphous, exhibiting liquid like organization of the constituents
(atoms, molecules or polymers),
and on the other hand is a solid, reacting elastically (reversibly) to small strains. There is a large variety of experimental examples of such glassy systems, and theoretically there are many well studied models \cite{07BK,ML,09LP} based on point particles with a variety of inter-particle potentials that exhibit stable supercooled liquids phases which then solidify to an amorphous solids when cooled below the glass transition. Typically all these materials, both in the lab and on the computer, begin to have plastic (irreversible) responses when the external strains increases beyond some limit. All these systems also exhibit a so-called yield-stress above which the material fails to a plastic flow, either homogeneously or by shear localization as seen in Fig. \ref{shearband}

\begin{figure}
\hskip -1.5 cm
\includegraphics[scale = 0.18]{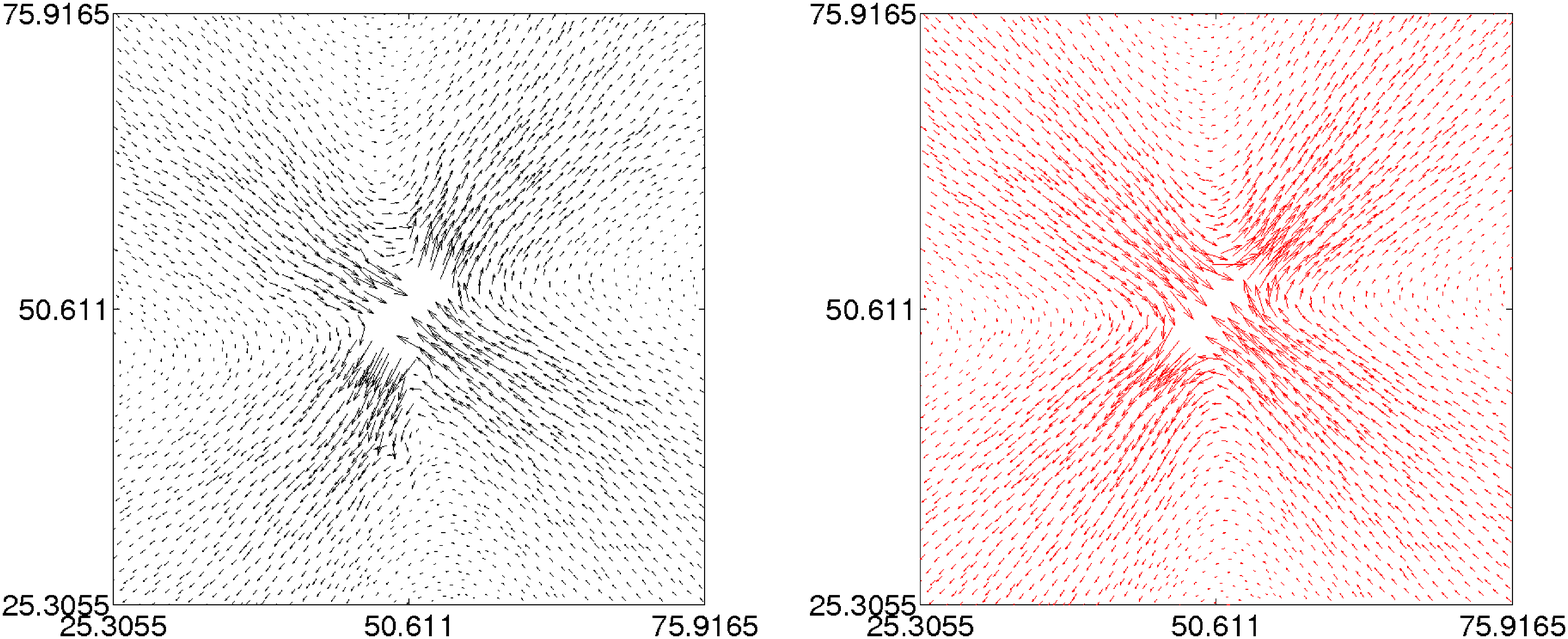}
\caption{(Color Online). Left panel: the localization of the non-affine displacement onto a quadrupolar structure which is modeled by an Eshelby inclusion, see right panel.  Right panel: the displacement field associated with a single Eshelby circular inclusion of radius $a$, see text.  The best fit parameters are $a \approx 3.4$ and $\epsilon^{*} \approx 0.09$ with a Poisson ration $\nu=0.363$. To remove the effect of boundary conditions, the best fit is generated on a smaller box of size $\left(x,y\right) \in \left[25.30 , 75.92\right]$. We find the parameters $a$ and $\epsilon^{*}$ to be very weakly dependent on the external strain for a given quench rate. }
\label{loc}
\end{figure}

To understand this phenomenon one needs to briefly review recent progress in understanding plasticity
in amorphous solids \cite{ML,09LP,10HKLP,10KLP,12DKP}. Below we deal with 2-dimensional systems composed of $N$ point particles in an area $A$, characterized by a total energy $U(\B r_1, \B r_2, \cdots \B r_n)$ where $\B r_i$ is the position of the $i$'th particle. Generalization to 3-dimensional systems is straightforward if somewhat technical. The fundamental plastic instability  is most cleanly described in athermal ($T=0$) and quasi-static (AQS) conditions when an amorphous solid is subjected to quasi-static strain, allowing to system to regain mechanical equilibrium after every differential stress increase. Higher temperatures and finite strain rates introduce fluctuations and lack of mechanical equilibrium which cloud the fundamental physics
of plastic instabilities with unnecessary details \cite{10HKLP}. The external strain is denoted below $\epsilon^\infty_{\alpha\beta}$
 with a shear component $\gamma\equiv\epsilon^\infty_{xy}$.  We choose to develop the theory for the case of external shear since then the strain tensor is traceless, simplifying some of the theoretical expressions. Applying an external shear, one discovers that the response of an amorphous solids to a small increase in the external shear strain $\delta\gamma$ is composed of two contributions. The first is the affine response which
simply follows the imposed shear, such that the particles positions $\B r_i={x_i,y_i}$ change via
\begin{eqnarray}
x_i &\to& x_i+\delta\gamma \,y_i \equiv x'_i\nonumber\\
y_i &\to& y_i \equiv y'_i.
\end{eqnarray}
 This affine response results in nonzero forces between the particles (in an amorphous solid) and these are relaxed by the non-affine response $\B u_i$ which returns
the system to mechanical equilibrium. Thus in total $\B r_i\to \B r'_i+\B u_i$. The nonaffine response $\B u_i$ solves an exact (and model independent) differential equation of the form \cite{ML,11HKLP}
\begin{equation}
\frac{d\B u_i}{d\gamma} = -H^{-1}_{ij} \Xi_j
\end{equation}
where $H_{ij} \equiv \frac{\partial^2 U(\B r_1, \B r_2, \cdots \B r_n)}{\partial \B r_i\partial \B r_j}$ is the so-called Hessian matrix and $\Xi_i\equiv
 \frac{\partial^2 U(\B r_1, \B r_2, \cdots \B r_n)}{\partial \gamma \partial \B r_i}$ is known as the non-affine force.
 The inverse of the Hessian matrix is evaluated after the removal of any Goldstone modes. A plastic event occurs when a nonzero eigenvalue $\lambda_P$ of $\B H$ tends to zero at some strain value $\gamma_P$. It was proven that this occurs universally via a saddle node bifurcation such that $\lambda_P$ tends to zero like $\lambda_P\sim \sqrt{\gamma_P-\gamma}$ \cite{12DKP}. For values of the stress which are below the yield stress the plastic instability is seen \cite{ML} as
 a localization of the eigenfunction of $\B H$ denoted as $\Psi_P$ which is associated with the eigenvalue $\lambda_P$, (see Fig. \ref{loc} left panel).
 While at $\gamma=0$ all the eigenfunctions associated with low-lying eigenvalues are delocalized, $\Psi_P$ localizes as $\gamma\to \gamma_P$
 (when $\lambda_P\to 0$) on a quadrupolar structure as seen in Fig. \ref{loc}for the non-affine displacement field. when the plastic instability is approached. These simple plastic instabilities involve the motion of a relatively small number of particles but the stress field that is released has a long tail.

 %%%%%%%%%%%%%%%%%%%%%%%%%%%%%%%%%%%%%%%%%%%%%%%%%%
%%%%%%%%%%%%%%%%%%%%%%%%%%%%%%%%%%%%%%%%%%%%%%%%%
%%%%%%%%%%%%%%%%%%%%%%%%%%%%%%%%%%%%%%%%%%%%%%%%%%
\begin{figure}
\hskip -2.0 cm
\includegraphics[scale = 0.25]{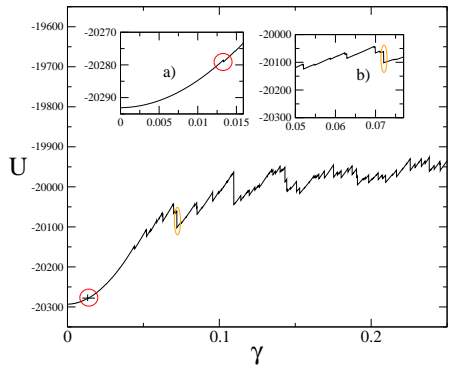}
\caption{(Color online). The energy vs. strain in a typical numerical simulation in our system. Shown are the points on the curve for a regular plastic event involving a single quadrupolar structure, (marked in a red circle on the main graph and in insert (a) which is a blow up of the same graph), and also the point on the curve that results in a plastic instability leading to a shear band, yellow ellipse on the main graph and the blow up in insert (b). The regular plastic event is not even seen without a blow up.  }
\label{example}
\end{figure}

\begin{figure}
%\hskip -1.5cm
\includegraphics[scale = 0.17]{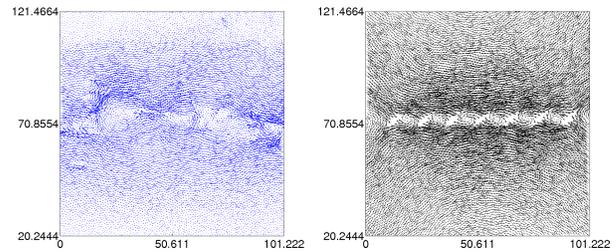}
\caption{(Color Online). Left panel: The nonaffine displacement field associated with a plastic instability that results in a shear band. Right panel: the displacement field associated with 7 Eshelby inclusions on a line with equal orientation. Note that in the left panel the quadrupoles are not precisely on a line as a result of the finite boundary conditions and the randomness. In the right panel the series of ${\C N}=7$ Eshelby inclusions, each given by Eq. (\ref{uc}) and separated by a distance of $13.158$, using the best fit parameters of Fig. \ref{loc}, have been superimposed to generate the displacement field shown. }
\label{result}
\end{figure}

 When the strain increases beyond some yield strain, the nature of the plastic instabilities can change in a fundamental way \cite{06TLB}. The main analytic calculation that is reported in this Letter shows that {\em when the stress built in the system is sufficiently large},
 instead of the eigenfunction localizing on a single quadrupolar structure, {\bf it can now localize on a series of $\C N$ such structures, which are organized on a line that is at $45$ degrees to the principal stress axis, with the quadrupolar structures having a fixed orientation relative to the applied shear}. In Fig. \ref{example} we show a typical stress vs. strain curve of a glassy material (the details of the simulations are presented in the accompanying material \cite{acc}), and show the non-affine displacement field associated with the plastic instability
 after exceeding the yield stress as indicated by the ellipse on the curve. Fig. \ref{result} shows the non-affine field that is identical to the eigenfunction which is associated with this instability, clearly demonstrating the series
 of quadrupolar structures that are now organizing the flow such as to localize the shear in a narrow strip around them. This is the fundamental shear banding instability.
Note that this instability is reminiscent of some chainlike structure seen in liquid crystals, arising from the orientational elastic energy of the anisotropic host fluid \cite{97PSLW}, and ferromagnetic
chains of particles in strong magnetic fields \cite{70GP}.

 To explain why this mechanism for shear banding can appear only at values of the stress that exceed
 the yield stress, we turn now to analysis. As a first step \cite{ML} we model the quadrupolar stress field which is associated with the simple plastic instability as a circular Eshelby inclusion of radius $a$ and a traceless eigenstrain
 $\epsilon^*_{\alpha\beta} = \epsilon^* (2\hat n_\alpha\hat n_\beta-\delta_{\alpha\beta})$ where $\hat{\B n}$ is a unit vector along one of the principal direction of the eigenstrain tensor \cite{57Esh}. The inclusion is inserted in a homogeneous elastic medium with Young modulus $E$ and Poisson ratio $\nu$ which is subjected to a homogeneous shear strain $\epsilon^\infty_{\alpha\beta}$. It was shown by Eshelby that inside the inclusion we have, due
  to the effect of the constraining elastic medium, a constant strain $\epsilon^c_{\alpha\beta} = S_{\alpha\beta\gamma\delta} \epsilon^*_{\gamma\delta}$ where $ S_{\alpha\beta\gamma\delta}$ is a constant tensor for any elliptical inclusion. For a circular inclusion the Eshelby tensor reads
 \begin{equation}\\
 S_{\alpha\beta\gamma\delta}=\frac{4\nu-1}{8(1-\nu)} \delta_{\alpha\beta}\delta_{\gamma\delta} +\frac{3-4\nu}{8(1-\nu)}(\delta_{\alpha\delta}\delta_{\beta\gamma}+\delta_{\beta\delta}\delta_{\alpha\gamma})
  .
 \end{equation}
Computing accordingly we find that the constrained strained is proportional to the eigenstrain, i.e. $\epsilon^c_{\alpha\beta} = \frac{3-4\nu}{4(1-\nu)} \epsilon^*_{\alpha\beta}$. {\em Inside} the inclusion, using the fact that a traceless strain field $\B \epsilon$ induces a stress field $\B \sigma = \frac{E}{1+\nu} \B \epsilon$ and displacement field $u_\alpha=\epsilon_{\alpha\beta}X_\beta$ (with $\B X$ denoting an arbitrary cartesian point in the material) we find
in the inclusion
\begin{eqnarray}
\sigma_{\alpha\beta}& = &\sigma_{\alpha\beta}^c -\sigma_{\alpha\beta}^* +\sigma^\infty_{\alpha\beta} =\frac{-E}{4(1-\nu^2)} \epsilon^*_{\alpha\beta}+\sigma^\infty_{\alpha\beta}\ , \\ u_\alpha&=& [\frac{3-4\nu}{4(1-\nu)} \epsilon^*_{\alpha\beta}+\epsilon_{\alpha\beta}^\infty] X_\beta \ .
\end{eqnarray}
Now {\em outside} of the inclusion the displacement field can be written as $u_\alpha(\B X)=u_\alpha^c(\B x) + \epsilon_{\alpha\beta}^\infty X_\beta$ where $u_\alpha^c(\B x)$ solves the bi-Laplacian equation
\begin{equation}
\nabla^2 \nabla^2 u^c_\alpha (\B X) =0 \ ,
\end{equation}
subject to continuity on the surface of the inclusion and zero at infinity. Remembering the radial solutions of the bi-Laplacian in 2-dimensions (i.e. $1, \ln r, r^2, r^2\ln r$), we write the most general displacement field that is linear in the traceless eigenstrain that tends to zero
at infinity:
\begin{eqnarray}\label{disp_field}
u_\alpha^c (\B X)&=&A  \epsilon^*_{\alpha\beta}\frac{\partial \ln r}{\partial X_\beta} +B  \epsilon^*_{\beta\gamma}\frac{\partial^3 \ln r}{\partial X_\alpha \partial X_\beta \partial X_\gamma} \nonumber\\
&+&C \epsilon^*_{\beta\gamma}\frac{\partial^3 (r^2\ln r)}{\partial X_\alpha \partial X_\beta \partial X_\gamma} \ ,
\end{eqnarray}
since the third derivative of $r^2$ vanishes identically. We determine the coefficients $A,B,C$ as usual by fitting the boundary conditions. The calculation is lengthy but standard (see a line-by-line solution in the accompanying material)
with the final result
\begin{eqnarray}
&&u_\alpha^c (\B X) =\label{uc}\\&&\frac{\epsilon^*}{4(1-\nu)}\left(\frac{a}{r}\right)^2\Big[2(1-2\nu) +\left(\frac{a}{r}\right)^2\Big]\Big[2\hat n_\alpha\B n\cdot\ \B X- X_\alpha\Big]\nonumber\\&&+\frac{\epsilon^*}{2(1-\nu)}\left(\frac{a}{r}\right)^2
\left[1-\Big(\frac{a}{r}\right)^2\Big]\Big[\frac{2(\B n\cdot\B X)^2}{r^2} -1\Big]X_\alpha \ . \nonumber
\end{eqnarray}

Having the displacement field associated with each Eshelby inclusion at hand, we can now turn to the
main calculation of the energy of $n$ such inclusions arranged at random positions in the material and with a random orientation of their quadrupole. Denoting the $n$ inclusions with an index $i=1,2,\cdots n$ and the externally induced stress field by $\sigma_{\alpha \beta}^\infty$ we can write the total energy of the material as $E=E^\infty + E_{\rm inc}+E_{\rm esh}+ E_{\rm mat}$.  The first is due to the
externally induced shear interacting with the strain field of the inclusions, the second is the interaction of the inclusions themselves (i.e. the stress of one with the strain of the other). The third is the self energy of the inclusions and the fourth the self energy of the strained material without
inclusions. For the purpose of this calculation we need only the first two:
\begin{eqnarray}
E^\infty &=& -\frac{\pi a^2}{2} \sigma^\infty_{\alpha\beta} \sum_{i=1}^n \epsilon^{(*,i)}_{\alpha\beta}\ , \label{E}\\
E_{\rm inc}&=&-\frac{\pi a^2}{2}\sum_{<ij>}\left[\epsilon^{(*,i)}_{\beta\alpha}\sigma^{(c,j)}_{\alpha\beta}(R_{ij})+
\epsilon^{(*,j)}_{\beta\alpha}\sigma^{(c,i)}_{\alpha\beta}(R_{ij})\right] \ . \nonumber
\end{eqnarray}
Here $R_{ij}$ is the distance between two inclusions and the stress field $\sigma^{(c,i)}_{\alpha\beta}(R_{ij})$  is that induced by inclusion $i$ at a distance $R_{ij}$ away. This stress can be readily computed from Eq. \ref{uc}. The computation of $E_{\rm inc}$ is very lengthy, and is reproduced in the supplementary
material where we also explain and justify the far-field approximation that is being used. The final result is
\begin{widetext}
\begin{eqnarray}
&&E_{\rm inc} \!=\! - \frac{\pi a^2(\epsilon^*)^2 E}{8(1-\nu^2)} \sum_{<ij>} (\frac{a}{R_{ij}})^2\Big\{
\!\!-8[(1-2\nu)+\!\!(\frac{a}{R_{ij}})^2][4\hat {\B n}^{(i)}\cdot \hat{ \B n}^{(j)} \hat {\B n}^{(i)}
\cdot \hat {\B r}_{ij} \hat {\B n}^{(j)} \cdot \hat {\B r}_{ij}\! -2\!(\hat {\B n}^{(i)}
\cdot \hat {\B r}_{ij})^2-2(\hat {\B n}^{(j)}
\cdot \hat {\B r}_{ij})^2+1] \nonumber\\
&&+4[2(1-2\nu) + (\frac{a}{R_{ij}})^2][2(\hat {\B n}^{(i)}\cdot \hat{ \B n}^{(j)})^2-1]
-8[1-2(\frac{a}{R_{ij}})^2][2(\hat {\B n}^{(i)} \cdot \hat {\B r}_{ij})^2-1][2(\hat {\B n}^{(j)} \cdot \hat {\B r}_{ij})^2-1] \label{Einc}\\
&&+32[1-(\frac{a}{R_{ij}})^2][\hat {\B n}^{(i)}
\cdot \hat {\B r}_{ij} \hat {\B n}^{(j)} \cdot \hat {\B r}_{ij}\hat {\B n}^{(i)}\cdot \hat{ \B n}^{(j)}-(\hat {\B n}^{(i)} \cdot \hat {\B r}_{ij})^2(\hat {\B n}^{(j)} \cdot \hat {\B r}_{ij})^2]\Big\}\ , \nonumber
\end{eqnarray}
\end{widetext}
where $\hat {\B r}_{ij}$ is the unit vector along the line connecting the $i$'th and $j$'th quadrupoles.
On the other hand we find
\begin{equation}
E^\infty = -\frac{\pi a^2 E \gamma \epsilon^*}{1+\nu} \sum_i^n n_x^{(i)} n_y^{(i)}  \ . \label{Einf}
\end{equation}
Our task is now to find the configuration of $\C N$ quadrupoles that minimize the total energy.
Obviously, if the external strain $\gamma$ is sufficiently large, we need to minimize
$E^\infty$ separately, since it is proportional to $\gamma$. The minimum of (\ref{Einf}) is
obtained for
\begin{equation}
n_x^{(i)}= n_y^{(i)} = \frac{1}{\sqrt{2}} \ .
\end{equation}
Substituting this in Eq. (\ref{Einc}) simplifies it considerably:
\begin{widetext}
\begin{eqnarray}
&&E_{\rm inc} = - \frac{\pi a^2(\epsilon^*)^2 E}{8(1-\nu^2)} \sum_{<ij>} (\frac{a}{R_{ij}})^2\Big\{
-8[(1-2\nu)+(\frac{a}{R_{ij}})^2]+4[2(1-2\nu) + (\frac{a}{R_{ij}})^2]
-8[1-2(\frac{a}{R_{ij}})^2][2(\hat {\B n} \cdot \hat {\B r}_{ij})^2-1]^2\nonumber\\
&&+32[1-(\frac{a}{R_{ij}})^2][(\hat {\B n} \cdot \hat {\B r}_{ij})^2-(\hat {\B n} \cdot \hat {\B r}_{ij})^4]
\Big\} \ . \label{final}
\end{eqnarray}
\end{widetext}
To find the minimum energy, denote $x_{ij}\equiv (\hat {\B n} \cdot \hat {\B r}_{ij})^2$, and minimize the sum by
minimizing in each term the expression $A[2x_{ij}-1]^2 - B[x_{ij}-x_{ij}^2]$. The minimum is obtained
at $x_{ij}=x=1/2$, meaning that all the $\hat{\B r}_{ij}=\hat{\B r}$, and this unit vector has an angle with $\cos \phi = \sqrt{1/2}$. We thus conclude that when the line of correlated quadrupoles forms under shear,  this line is in 45 degrees to the compressive axis, as is indeed
seen in experiments. 

The physical meaning of this analytic result is that it is cheaper (in energy) for the material
to organize $\C N$ quadrupolar structures on a line of 45 degrees with the compressive stress, all having the same orientation, than any other arrangement of these $\C N$ quadrupoles, including any
random distribution. This explains why such a highly correlated distribution appears in the strained
amorphous solid, and why it can only appear when the external strain (or the built-up stress)
are high enough. This fact, in addition to the observation that such an arrangement of Eshelby quadrupoles localizes the shear, explains the origin of this fundamental instability.

It should be noted that in the present calculation we did not {\em predict} the number of quadrupoles that appear at the instability. To achieve this we must consider the other terms in the total energy (i.e. $E_{\rm esh}$ and $E_{\rm mat}$ above), and this is beyond the scope of this Letter. This calculation
and the resulting theoretical estimate of the yield stress will be presented in later publications.
For example one can show that the yield strain is a function of the parameter $\epsilon^*$, in the form $\gamma_Y\equiv \epsilon^*/[2(1-\nu)]$ \cite{12DHP}. $\epsilon^*$ is controlled to some degree by the protocol of quench from the melt to the solid \cite{12DHP}.

Acknowledgements: This work had been supported in part by the Israel Science Foundation, the German-Israeli Foundation and by the European Research Council under an ``ideas" grant.

\end{document}